\documentclass{article}

\usepackage{arxiv}

\usepackage[utf8]{inputenc} 
\usepackage[T1]{fontenc}    
\usepackage{hyperref}       
\usepackage{url}            
\usepackage{booktabs}       
\usepackage{amsfonts}       
\usepackage{nicefrac}       
\usepackage{microtype}      
\usepackage{lipsum}
\usepackage{graphicx}
\usepackage{amsthm,amsmath}
\usepackage{etoolbox}

\title{Information Access Equality on \\ Network Generative Models}

\author{
 Xindi Wang \\
  Network Science Institute\\
  Northeastern University\\
  Boston, MA \\
  \texttt{wang.xind@northeastern.edu} \\
   \And
 Onur Varol \\
  Faculty of Engineering and Natural Sciences\\
  Sabancı University\\
  Istanbul, Turkey\\
  \texttt{onur.varol@sabanciuniv.edu} \\
  \And
 Tina Eliassi-Rad \\
  Network Science Institute\\
  Khoury College of Computer Sciences\\
  Northeastern University\\
  Boston, MA \\
  \texttt{tina@eliassi.org} \\
}

\begin{document}
\maketitle
\begin{abstract}
It is well known that networks generated by common mechanisms such as preferential attachment and homophily can disadvantage the minority group by limiting their ability to establish links with the majority group. This has the effect of limiting minority nodes' access to information. We present the results of an empirical study on the equality of information access in network models with different growth mechanisms and spreading processes. For growth mechanisms, we focus on the majority/minority dichotomy, homophily, preferential attachment,  and diversity.  For spreading processes, we investigate simple vs.~complex contagions, different transmission rates within and between groups, and various seeding conditions. We observe two phenomena. First, information access equality is a complex interplay between network structures and the spreading processes. Second, there is a trade-off between equality and efficiency of information access under certain circumstances (e.g., when inter-group edges are low and information transmits asymmetrically).  Our findings can be used to make recommendations for mechanistic design of social networks with information access equality.
\end{abstract}

\section{Introduction}
The early hopes that an increasingly interconnected world will provide greater equality of opportunity and remove information barriers are tempered by observations of echo chambers and polarisation on social networks~\cite{adamic2005political,iyengar2009red,prior2007post,garrett2009echo,gentzkow2011ideological,barbera2015tweeting}. Recent studies have also found that discrimination in social networks can arise through simple mechanisms, for example, homophily and minority group size can lead to a glass-ceiling effect~\cite{avin2015homophily,stoica2018algorithmic}, a chasm effect~\cite{zhang2021chasm}, influence the ranking of minority nodes~\cite{karimi2018homophily,oliveira2021mixing}, and create perception bias in social networks~\cite{lee2019homophily}.
Furthermore, a user's popularity and prestige on social networking platforms can be altered through the use of automation and social bots to gain unfair advantages.~\cite{aiello2012people,messias2013you,woolley2016automating,varol2017online,varol2020journalists}.

An important role of networks is that they disseminate information, from job and business opportunities~\cite{boyd2014networked} to medical resources that can be accessed~\cite{freese2011fundamental}, and such information is critical to people's lives. However, inequality in access to information in a network has not been fully explored. This is a topic that has been discussed in the social and political sciences, with examples such as economically disadvantaged people having less access to new technologies such as the Internet Internet~\cite{dimaggio2021information}, and groups starting out with poorer employment status being more likely to experience persistent unemployment in the labour market~\cite{calvo2004effects}. 

A closely related research topic to information dissemination is information maximization, which aims to find optimal starting (or seeding) nodes in a network that maximize information spreading~\cite{kempe2003maximizing}. This problem has been studied over the years, but the fairness aspect in this problem is relatively new. For example, Stoica and Chaintreau~\cite{stoica2019fairness} investigate two fairness criteria for information maximization: fairness for early adopters (where seeds should be proportional to the group population) and fairness in outreach (where final outreach nodes should reflect the group population). They experiment on a social network extracted from Instagram. In subsequent work, Stoica et al.~\cite{stoica2020seeding} proposed diversity-enhancing interventions in seeding and explored the complicated relationship between diversity and efficiency. Other research explored how fairness constraints can be incorporated into the information maximization algorithm, including the maximin constraint (which requires that the least wealthy group should be improved during the optimization process)~\cite{fish2019gaps,becker2020fairness}, diversity constraint~\cite{tsang2019group}, welfare theory~\cite{rahmattalabi2020fair}, and adding a time-sensitive constraint on information~\cite{ali2019fairness}. These papers shed some initial light on information equality in networks, but they mostly consider the problem as an algorithmic problem (given a fixed graph, what are the optimal information seeds), rather than a characteristic problem (what kind of network can better promote information equality). Moreover, the spreading process they consider is rather limited. Most of the work is concerned with studying the information cascade model, which may not be representative of real-world processes. Different from previous works, Jalali et al.~\cite{jalali2020information} define a criterion for information unfairness (based on whether information flows equally among all groups in a network) and present an algorithm that adds edges to the network to reduce the information unfairness criterion. They experiment on a social network extracted from DBLP. This study does not provide much information about the performance of various spreading processes on different networks. In 2021, Venkatasubramanian et al.~\cite{kdd21tutorial} produced a much-needed tutorial on fairness in networks. Their tutorial covers social capital, information access, and interventions.

Here, we study the problem of information access equality in networks from a characteristic point of view. Using several network models with two mutually exclusive groups in the population (majority vs.~minority), we generate networks with different properties and constraints that are representative of mechanisms in real networks. We define information access equality as follows: \emph{for a given process and seeds, the majority and minority nodes should receive information at similar rates across various stages of the spreading processes}. Our goal is to provide insight into which network characteristics may affect information access equality and in what ways, and to make recommendations for systematic mechanism design. We find that, in general, homophily and preferential attachment can harm information access equality, while introducing diversity can promote information access equality. However, too much diversity can affect the efficiency of information spreading. We also find that information access equality depends not only on the network, but also on the characteristics of the spreading process.

\section{Network Models}
Our study includes several generative models for complex networks. Each model generates a network with two groups of nodes: majority and minority. The proportion of minority nodes is $m$. The majority/minority dichotomy is based on population size: minority nodes make up less than 50\% of the network (i.e. $m < 0.5$). A real-world example of such a dichotomy is a computer science collaboration network where male scientists are in the majority and female scientists are in the minority.

The generative models produce undirected, unweighted networks. Starting from an initialized network, a new node enters the network at each time step and connects to some nodes according to a mechanism mandated by the model. The initialization procedure is the same across the models: a node from the majority group connects to a node from the minority group. Table~\ref{tab:model} summarizes the properties of each model and provides comparison between the proposed models.

\begin{table}[!ht]
\centering
\footnotesize
\begin{tabular}{ccccc}
\textbf{Random Network} & \begin{tabular}[c]{@{}c@{}}\textbf{Preferential} \\ \textbf{Attachment}\end{tabular} & \textbf{Homophily} & \textbf{Diversity} & \textbf{Resultant Network}           \\ \hline\hline
\checkmark  & \checkmark    &           &           & BA \\ \hline
\checkmark &     & \checkmark &           & Random Homophily  \\ \hline
\checkmark &  \checkmark &  \checkmark  &   & Homophily BA  \\ \hline
\checkmark  &   & \checkmark  &  \checkmark & \begin{tabular}[c]{@{}c@{}}Diversified \\ Homophily\end{tabular}   \\ \hline
\checkmark &  \checkmark &  \checkmark  &  \checkmark  & \begin{tabular}[c]{@{}c@{}}Diversified \\ Homophily BA\end{tabular} \\ \hline
\end{tabular}
\caption{\textbf{Relationship between the Network Models.} All network models are based on Random Network, a growing network adaptation of the Erd\H{o}s-R\'{e}nyi random network~\cite{erdHos1960evolution}. We focus on three mechanisms: preferential attachment, homophily, and diversity (through adding inter-group edges). Starting from a Random Network and adding preferential attachment results in the BA model. Starting from a Random Network and adding homophily results in the Random Homophily model. Starting from a Random Network and adding both preferential attachment and homophily results in the Homophily BA model. Starting from Random Homophily and adding diversity results in the Diversified Homophily model. Starting from Homophily BA and adding diversity gives us Diversified Homophily BA model. In our study, we treat BA and Random Homophily as the variations of the Homophily BA model, and Diversified Homophily as a variation of the Diversified Homophily BA. \label{tab:model}}
\end{table}

\subsection*{Random Network}
In the Random Network model, at each time step a new node enters with probability $m$ as the minority and $1-m$ as the majority. With uniform probability, each node connects to $l$ existing nodes. This is a growing version of the Erd\H{o}s-R\'{e}nyi Random Network~\cite{erdHos1960evolution}.

\subsection*{Homophily BA}
In the Homophily BA model~\cite{karimi2018homophily,lee2019homophily},\footnote{BA is short for Barab\'{a}si-Albert.} there are two ingredients in link formation: preferential attachment and homophily. The parameters of Homophily BA include: (1) the proportion of minority nodes $m$, (2) the number of edges $l$ for each new node, (3) the homophily matrix $H$ with entries $h_{g_i g_j}$,  and (4) the preferential attachment strength $\alpha$. $g_i$ and $g_j$ denote the group memberships for nodes $i$ and $j$, respectively. Each node $i$ can either belong to the majority (maj) or the minority (min) group. Thus, we have the following homophily matrix $H$:
$$
H = \begin{bmatrix}
h_{maj, maj} & h_{maj, min}\\
h_{min, maj} & h_{min, min}
\end{bmatrix}
$$
For simplicity, we assume $H$ is a doubly stochastic matrix, with $h_{maj, min} = h_{min, maj}$ and $h_{maj, maj} = h_{min, min}$. We define $h = h_{maj, maj} = h_{min, min}$. When $h = 1$, the network is perfectly homophilic, i.e., the network has two distinct groups: majority and minority. The groups are connected with a single inter-group edge (due to network initialization).  On the other hand, if $h = 0$, the network is perfectly heterophilic. It has been found that homophily can be different for different groups~\cite{messias2017white, karimi2018homophily}. We leave the exploration of asymmetric homophily for future work. 

The Homophily BA networks grow as follows. 

\begin{itemize}
    \item At each time step, a new node $j$ enters with probability $m$ as the minority and $1-m$ as the majority. Denote its group as $g_j$.
    
    \item Node $j$ connects to $l$ nodes. Each connection to node $i$ in the network is made with probability $\Pi_i$:
    \begin{equation}
        \Pi_i = \frac{h_{g_j g_i} d_i^\alpha}{\sum_i{h_{g_j g_i} d_i^\alpha}},
    \end{equation}
    where $d_i$ is the degree of node $i$.
\end{itemize}

We are also interested in two special cases of Homophily BA: (1) without preferential attachment (Random Homophily with $\alpha = 0$) and (2) with random mixing (BA~\cite{albert2002statistical} with $h = 0.5$).

\subsection*{Diversified Homophily BA}
We propose the Diversified Homophily BA model to encourage inter-group connections while maintaining some degree of homophily. The parameters of the Diversified Homophily BA include: (1) the proportion of minority nodes $m$, (2) the number of edges $l$ for each new node, (3) the homophily matrix $H$ (with the same assumption as in Homophily BA), (4) preferential attachment strength $\alpha$, (5) the number of diversified edges for each node $l_d$, and (6) the diversification probability $p_d$. The Diversified Homophily BA networks grow as follows:

\begin{itemize}
   \item  At each time step, a new node $j$ enters with probability $m$ as the minority and $1-m$ as the majority. Denote its group as $g_j$.
    
    \item Node $j$ forms $l-l_d$ links following Homophily BA mechanism 
    $\Pi_i = \frac{h_{g_j g_i} k_i^\alpha}{\sum_i{h_{g_j g_i} k_i^\alpha}}$. Denote the nodes connected at this step as $S_j$.
    
    \item Node $j$ forms $l_d$ diversified links. Given diversification probability $p_d$, we have the connecting probability of two nodes $j$ and $k$ as 
    \begin{equation}
    p_{jk} =
    \begin{cases}
        p_d, & {g_j \neq g_k}\\
        1 - p_d, & {g_j = g_k}.
    \end{cases}
    \end{equation}
   For node $i \in S_j$, we obtain their neighbors, denoted as $N_{S_j}$. Generate $l_d$ links by connecting to $k \in N_{S_j}$ with probability $\Pi_{jk} \propto p_{jk} \times \frac{1}{|d_k - d_i|}$, where $d_i$ is the degree of node $i \in S_j$. The idea behind this step is to connect to nodes which are of opposite group, but with similar degree to existing neighbors.
\end{itemize}

We are interested in a special case under Diversified Homophily BA with $\alpha = 0$, which removes preferential attachment. We call this case Diversified Homophily.

\section{Information Access Equality}
There is currently no consensus on how to measure information access equality in a network. In this work, we choose to measure it experimentally, i.e., by simulating spreading processes on the network and observing the differences in spreading among majority and minority nodes. This allows us to explore more possibilities in the process configurations, such as the type of contagion, the transmission rate, and the information seeds.

\subsection*{Varying Processes that Spread Information}
We consider several variations of dynamic processes. The first variation we consider is the distinction between \textit{simple} vs.~\textit{complex contagion}. Studies have found that while simple contagion is appropriate for most diseases and the spread of information, complex contagion is more common for collective behaviors, such as the spread of new technologies and innovations, the growth of social movements, spread of misinformation, etc.~\cite{centola2007complex,vespignani2012modelling,monsted2017evidence,granovetter1978threshold,centola2010spread,anderson1992infectious,daley1964epidemics,romero2011differences,weng2013virality,shao2018spread}. Same network properties may respond differently to these two types of contagion. For example, in simple contagion, weak ties are considered important because they can disseminate information to an isolated part of the network. However, in complex contagion, weak ties might not be as useful because one link is not sufficient to disseminate information~\cite{granovetter1973strength,centola2007complex}. For modeling, we rely on a Susceptible-Infectious (SI) model~\cite{barrat2008dynamical} for both simple and complex contagion. For complex contagion, we add the parameter of activation threshold $a$, where each node can only be infected if $a$ portion of its neighborhood is infected.

The second variation we consider is the group difference in information \textit{transmission rate}. Although the transmission rate can be different at the node level~\cite{aral2018social},  we study the simple case where the transmission rate differs by group. More specifically, we are interested in the case where the transmission rate between all nodes is the same (symmetric) and the case where the transmission rate between different groups is lower than within the same group (asymmetric). There are four sets of transmission rates: $r_{maj\rightarrow maj}$, $r_{min\rightarrow min}$, $r_{maj\rightarrow min}$ and $r_{min\rightarrow maj}$. Here we assume that $r_{maj\rightarrow maj}=r_{min\rightarrow min}=r_{within}$ and $r_{maj\rightarrow min} = r_{min\rightarrow maj} = r_{between}$.  We simulate the SI process with symmetric transmission rate as $r_{within} = r_{between}$, and asymmetric transmission rate as $r_{within}> r_{between}$.

The last variation we consider is the location where the information is seeded. Different seeding locations can advantage or disadvantage a group. For example, imagine a network with two groups (majority vs.~minority) connected by a single edge. Under this circumstance, if all seeds belong to the majority community, we would expect the minority group to be disadvantaged. To test the influence of seeding, we assume random seeding (as opposed to targeted seeding, such as selecting higher degree nodes). We assume different portion of minority seeds: low (minority seeds below 30\%), mid (minority seeds between 30\% and 70\%), and high (minority seeds above 70\%). For brevity, we show only the low and high minority seeding portions. The results for mid and high minority seeding portions are similar.

\subsection*{Measuring Information Access Equality}
Since spreading processes can take different times in different networks, we first normalize the fraction of nodes in state I (Infected) at each time step $I(t)$ by the length of the spreading process T to obtain $I(t/T)$. Then, we separate majority ($I_{maj}(t/T)$) and minority ($I_{min}(t/T)$) groups and calculate the relative difference between them as
$$\Delta I(t/T) = \frac{I_{maj}(t/T) - I_{min}(t/T)}{I_{maj}(t/T) + I_{min}(t/T)}.$$
We choose the denominator to be $I_{min}(t/T) + I_{maj}(t/T)$, which corresponds to twice the mean of $I_{maj}(t/T)$ and $I_{min}(t/T)$, making this metric symmetric and bounded between -1 and 1. When there is information access equality, $\Delta I(t/T) = 0$; and the closer $\Delta I(t/T)$ is to 0, the greater the equality between the two groups. $\Delta I(t/T) >0$ means that the majority group has a greater advantage and $\Delta I(t/T) <0$ means that the minority group has a greater advantage. This measure is intuitive and it allows us to inspect information access equality at different stages of the spreading process.

\section*{Network Measures}
To understand the structure and properties of the network, and to gain insight into possible roots of information access inequality, we examine several network measures. The first set of measures is dyadicity and heterophilicity~\cite{park2007distribution}. The second set of measures are regarding the distance in the network, which is directly related to information spreading. And the last set of measures are regarding degree inequalities between the two groups, as a measure of difference in social capital~\cite{burt2000network, berlingerio2012asonam}.

\subsection*{Homophily: Dyadicity and Heterophilicity}
The homophily parameter $h$ in Homophily BA and Diversified Homophily BA may not reflect the actual homophily level of the final network. To measure the homophily effect more accurately, we calculate the dyadicity and heterophilicity score of the network~\cite{park2007distribution}. For a network with random mixing of group membership, dyadicity and heterophilicity are expected to be close to 1.

\subsection*{Distance: Average Shortest Path Length and Diameter}
We measure the average shortest path length and the diameter of the network. These two measures are directly related to the spreading efficiency. The higher the shortest path length or diameter, the longer it takes for information to spread to the entire network.

\subsection*{Social Capital: Degree Equality}
We are interested in whether there is a relationship between equality of social capital and equality of information access. We use several measures of degree equality.

\paragraph{Earth Mover Distance (EMD).} We calculate the Earth Mover Distance (EMD) between two distributions. The smaller the distance, the more equal the two groups are.

\paragraph{Power Inequality (PI).} Power inequality~\cite{avin2015homophily} measures the ``power" (average degree) ratio of the minority and majority groups. Let $\bar{d}_{min}$ and $\bar{d}_{maj}$ denote the average degrees of the minority and majority nodes, respectively. The power inequality is defined as $PI = \frac{\bar{d}_{min}}{\bar{d}_{maj}}$. A network with power equality should have $PI = 1$; and the lower the $PI$, the more disadvantaged are minority nodes.

\paragraph{Moment Glass Ceiling (g).} Moment glass ceiling measures the glass-ceiling effect in a network~\cite{hymowitz1986glass, avin2015homophily}. The intuition behind it is that a larger second moment (and assuming a similar average degree, i.e., no power inequality) leads to a larger variance in the distribution and thus a significantly larger number of nodes with high degree (i.e., hubs). Let $E(d_{min}^2)$ and $E(d_{maj}^2)$ denote the second moment of the degree distribution for the minority and majority nodes, respectively. The moment glass ceiling is defined as $g = \frac{E(d_{min}^2)}{E(d_{maj}^2)}$. A network with no glass-ceiling effect should have $g = 1$; and the lower the $g$, the more disadvantaged are minority nodes.

\section{Experiments}
For simplicity, we control the parameter space for the information spreading simulations. For all synthetic models, we set the number of seed nodes to be  $s = 10$. In symmetric transmission, $r_{within} = r_{between} = 0.7$. In asymmetric transmission, $r_{within} = 0.7$ and $r_{between} = 0.3$. The activation threshold for complex contagion is $a = 0.1$.

\subsection*{Experiment 1: Information Access Equality across Different Network Models}
We create networks with $N = 5000$ nodes, $m=0.2$ (i.e., 20\% of the nodes are minority nodes) and $l = 2$. We set $h = 0.8$ and $\alpha = 1$ in Homophily BA and Diversified Homophily BA, including their variations (namely, Random Homophily and Diversified BA). For Diversified Homophily BA, we set $l_d = 1$ and $p_d=0.6$.

\subsubsection*{Network Measures}
\begin{figure}[!h]
    \centering
    \includegraphics[scale=0.36]{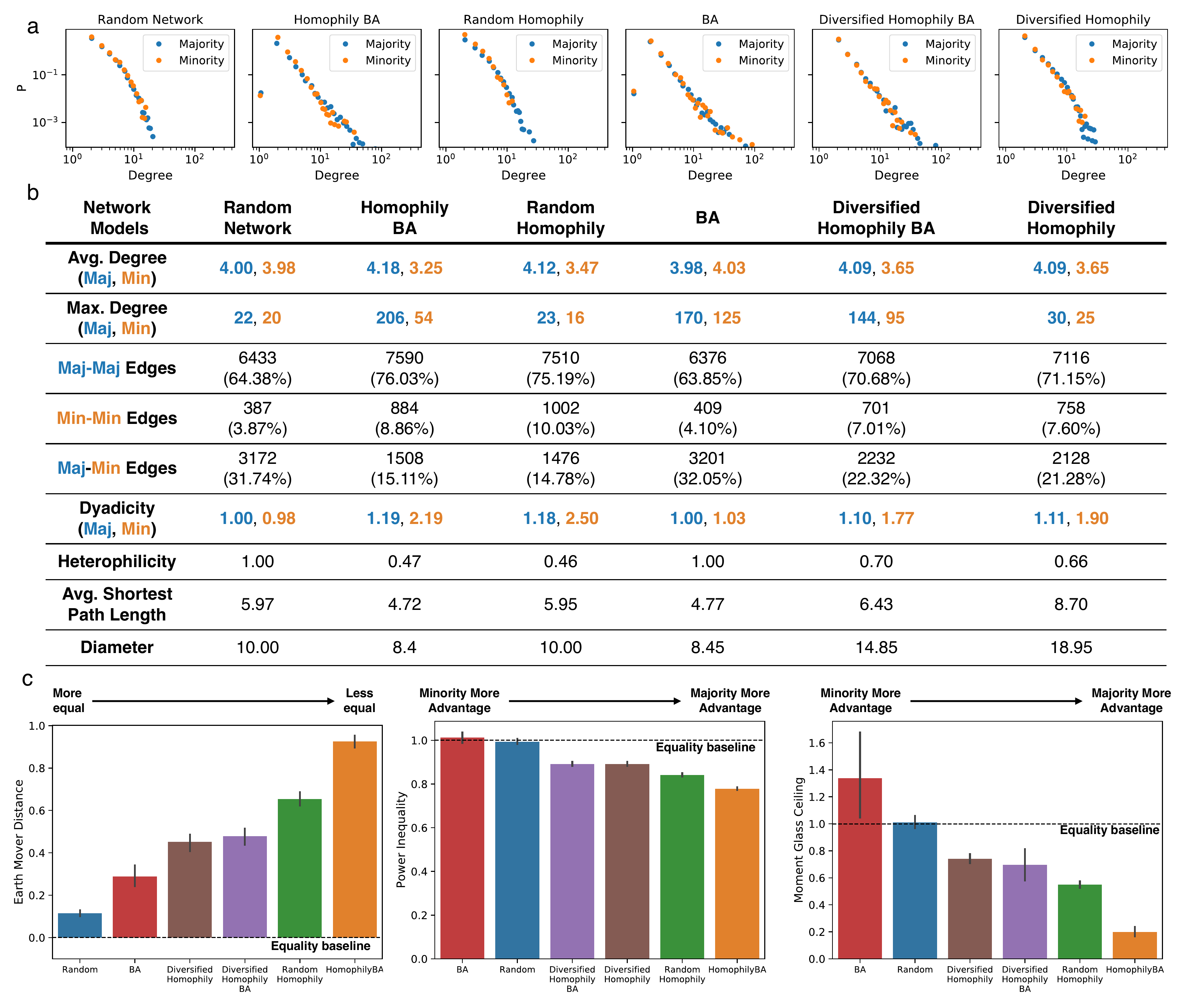}
    \caption{\textbf{Statistics of Network Models.} All networks have $N = 5000$ nodes, where each node joins the network with $l = 2$ edges and 20\% of the nodes are minority. We set $h = 0.8$, $e_o = 1$, $\alpha = 1$ and $p_d = 0.6$. \textbf{Degree distributions (a)} and \textbf{Basic Statistics (b).}
 Compared to Random Network, we see that Homophily BA has a longer tail in the degree distribution, a larger gap in average and maximum degree between the two groups, and a much smaller proportion of majority-minority edges, leading to an increase in dyadicity and a decrease in heterophilicity. Random Homophily decreases the portion of majority-minority edges more than Homophily BA, but it has a smaller gap in degree for the two groups, resulting from the absence of Preferential Attachment, leading to heavy-tail degree distribution. 
    BA model has a larger gap in maximum degree differences between the two groups compared to Random Network, but smaller than Homophily BA. The edge composition is similar to Random Network.
 Compared to Homophily BA, Diversified Homophily BA has a smaller gap in the degrees and also a higher portion of majority-minority edges, resulting in a higher heterophilicity score. 
 A similar phenomenon is observed in Diversified Homophily but with a smaller maximum degree for both groups due to the absence of Preferential Attachment. Looking at the average shortest path length and diameter, we find that Diversified Homophily BA and its variation have significantly higher average shortest path length and diameter, while Homophily BA and BA have lower average shortest path length and diameter.
 \textbf{(c) Social Captial Equality Measures.} For each equality measure, we sort the generative models from most equal (left) to least equal (right). The earth mover distance is the distance between the degree distributions of the minority and majority groups. The power inequality is the ratio of the average degree in the minority group to the average degree in the majority group. The moment glass ceiling is the ratio of the second moment of degree distribution in the minority group to the second moment of degree distribution in the majority group.  Across all three equality measures, from more equal to less equal, the rankings are Random Network or BA, followed by Diversified Homophily or Diversified Homophily BA, and lastly Random Homophily or Homophily BA.}
        \label{fig:network_stat}
\end{figure}

In Fig.~\ref{fig:network_stat}a and Fig.~\ref{fig:network_stat}b, we compare the networks generated by the various models based on their degree distributions and statistics on degrees, edges, dyadicity, heterophilicity, and average shortest path and diameter. For the degree distribution, we show the result of one sample network under each model. For all other statistics, we show the average of 20 realizations.

As expected, we observe that Random Network has dyadicity and heterophilicity of 1, and BA is similar. For the other models, we note that the dyadicity for both groups is above 1 and the heterophilicity is below 1, suggesting that there are fewer edges between the groups than in random mixing. We also note that, except for the Random Network, the dyadicity in the minority group is larger than the dyadicity in the majority group. Homophily BA and its variation have higher dyadicity and lower heterophilicity compared to Diversified Homophily BA and its variation. This observation makes sense since Diversified Homophily BA intentionally introduces inter-group edges.

We observe that Homophily BA and BA have lower average shortest path lengths and diameters due to the presence of hubs. Diversified Homophily BA has a slightly higher average shortest path length and diameter than Random Network and Random Homophily, while Diversified Homophily has even higher values, possibly due to the elimination of preferential attachment.

Figure~\ref{fig:network_stat}c shows the social captial (degree) equality measures across  different models.  For Earth Mover Distance, Power Inequality, and Moment Glass Ceiling, the rankings of the models from more equality to less equality is Random Network or BA, followed by Diversified Homophily or Diversified Homophily BA, and lastly Random Homophily and Homophily BA.

\subsubsection*{Information Access Equality}
\begin{figure}[ht!]
    \centering
    \includegraphics[scale = 0.38]{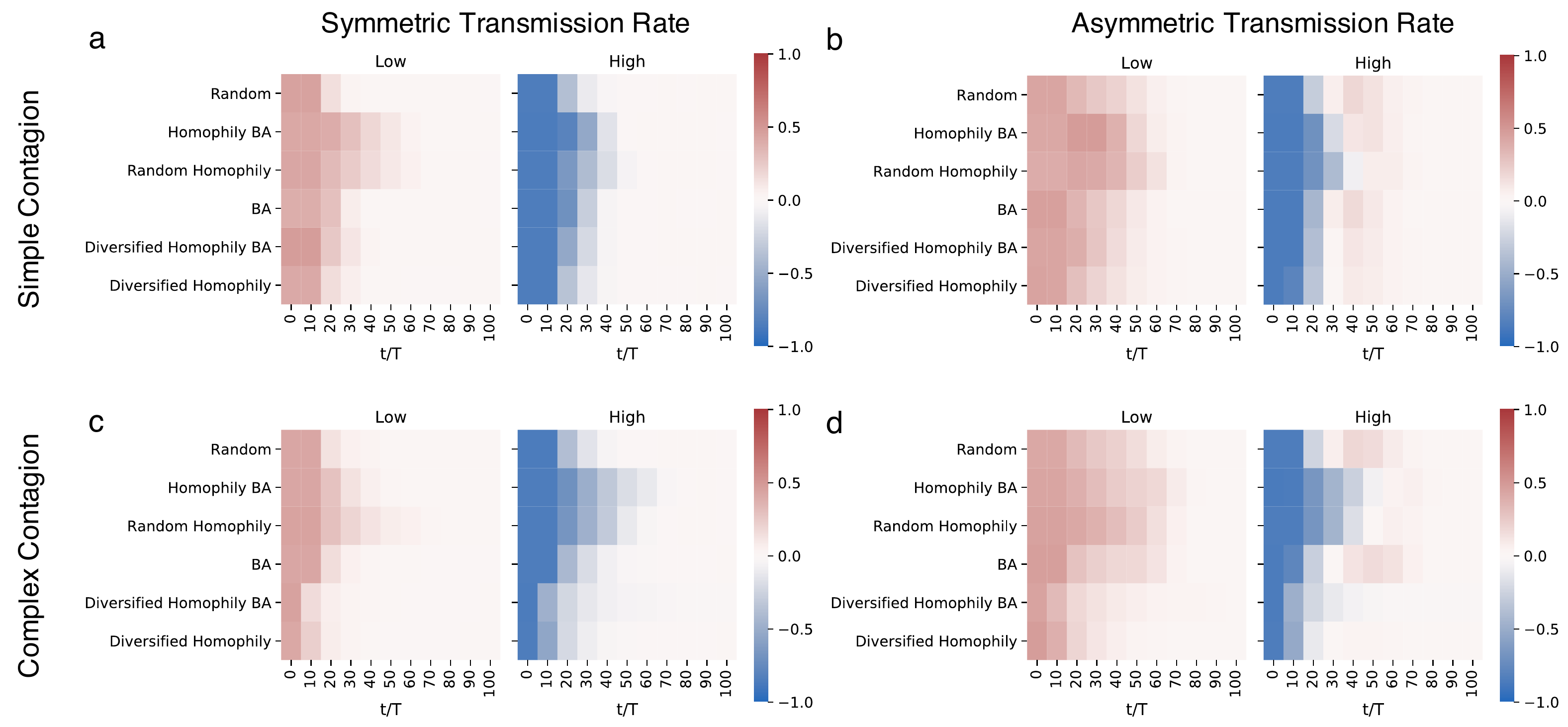}
    \caption{\textbf{Information Spreading Equality of Different Network Models.} We plot the spreading equality with low and high minority seeding portions. Each plot is a heatmap, where the x-axis represents the relative time $t/T$, the y-axis represents the different models, and the color represents $\Delta I(t/T)$. Recall that $\Delta I(t/T) = 0$ represents equality. When the minority seeding portion is low, $\Delta I(t/T)$ is initially positive (i.e., the majority group has a greater advantage); it then decreases to 0 for all models. When the minority seeding portion is high, $\Delta I(t/T)$ is initially negative (i.e., the minority group has a greater advantage). Under symmetric transmission, $\Delta I(t/T)$ increases to 0. However under asymmetric transmission, $\Delta I(t/T)$ increases to 0, then becomes positive, and eventually becomes 0. We observe that regardless of the contagion type and the seeding condition, Homophily BA and Random Homophily take longer to reach $\Delta I(t/T) = 0$, indicating less equality. On the contrary, Diversified Homophily BA and Diversified Homophily reach $\Delta I(t/T) = 0$ faster, sometimes even faster than Random Network and BA, indicating more equality. The differences between the models are more pronounced under complex contagion. 
     }
        \label{fig:network_info_equality}
\end{figure}

In Fig.~\ref{fig:network_info_equality}, we plot information access equality as a heatmap, where the x-axis represents the different stages of the process $t/T$, and the y-axis is the different models.  The color of each cell shows $\Delta I(t/T)$. $\Delta I(t/T) = 0$ is colored white (denotes equality); red color means $\Delta I(t/T) >0$ and the majority group is at an advantage, while blue color means $\Delta I(t/T) <0$ and the minority group is at an advantage. We omit the performances under mid minority seeding because we found they are very close to performance under high seeding portion.

Under low minority seeding portion, the majority group initially has an advantage ($\Delta I(t/T) >0$) for all models, and subsequently converges to equality ($\Delta I(t/T) = 0$). Comparing the models, we observe that Diversified Homophily BA and its variation take shorter time to reach equality, while Homophily BA and Random Homophily take longer time. When the minority seeding portion is high, the minority group initially has an advantage ($\Delta I(t/T) < 0$) and then converges to equality ($\Delta I(t/T) = 0$). Under asymmetric transmission, majority group has an advantage ($\Delta I(t/T) > 0$) in the middle of the process for some models. We again observe that Diversified Homophily BA and its variation take shorter time to reach equality, while Homophily BA and Random Homophily take longer time.

In summary, Homophily BA and Random Homophily achieve the lowest information access equality, followed by Random Network, BA and Diversified Homophily BA and their variation. The ranking between Random Network, BA, and Diversified Homophily BA depends on the process. Although Random Network and BA are most equal in degree equality measures and have the highest heterophilicity score, they are not always the most equal in information access equality (e.g., under complex contagion and asymmetric transmission rate).

\subsection*{Experiment 2: The Effects of Different Network Parameters on Information Access Equality}
We investigate the information access equality of network models with different parameters. Specifically, we investigate Homophily BA under different homophily $h$, preferential attachment strength $\alpha$, and minority portion $m$. We also investigate Diversified Homophily BA under different diversity $p_d$.

\begin{figure}[ht!]
    \centering
    \includegraphics[scale = 0.38]{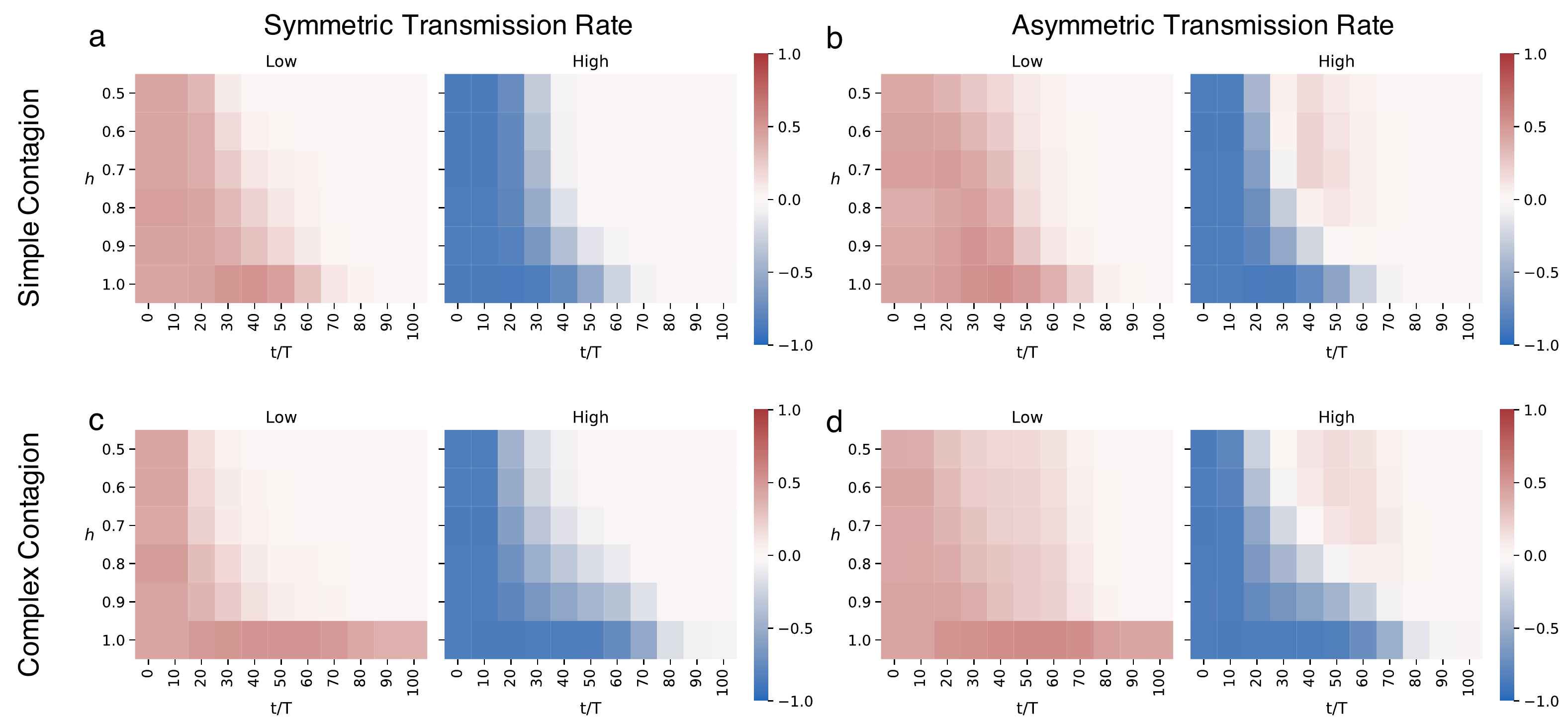}
    \caption{\textbf{Information Spreading Equality for Homophily BA with Different Levels of  Homophily $h$.} 
    We plot the spreading equality for Homophly BA networks with low and high minority seeding portions. Each plot is a heatmap, where the x-axis represents the relative time $t/T$, the y-axis represents the graph's homophily $h$, and the color represents $\Delta I(t/T)$. Recall that $\Delta I(t/T) = 0$ represents equality. At low minority seeding, the lower $h$ values, the more information access equality. This holds across all processes. However at high minority seeding portion, the behavior depends on the symmetry of the transmission. Under symmetric transmission rates (shown in (a) and (c)) with high minority seeding portion, lower $h$ values achieve more information access equality. Under asymmetric transmission rates (shown in (b) and (d)) with high minority seeding portion, lower $h$ values increase to zero, then become positive, before finally decreasing to zero. This makes it hard to judge which $h$ value achieves more equality when one has an asymmetric transmission and high minority seeding.
     }
     \label{fig:homoBA_h_info_equality}
\end{figure}

\paragraph{What is the influence of homophily on information access equality?} We first investigate the impact of homophily on information access equality (see Fig.~\ref{fig:homoBA_h_info_equality}). We vary homophily $h$ in the range between 0.5 (random mixing) and 1 (perfectly homophilic) under Homophily BA, setting $m = 0.2$, $l = 2$, $\alpha = 1$. At low minority seeding portion, we observe that for all processes lower $h$ values have more information access equality. At high minority seeding portion, under symmetric transmission rate (Fig.~\ref{fig:homoBA_h_info_equality}a, c), lower $h$ values also achieve more information access equality. However, under asymmetric transmission rate and high minority seeding portion, (Fig.~\ref{fig:homoBA_h_info_equality}b, d), we observe that the majority nodes gain an advantage around $t/T = 40$, keep that advantage until $t/T = 70$ before equality is established. 

\begin{figure}[ht!]
    \centering
    \includegraphics[scale = 0.38]{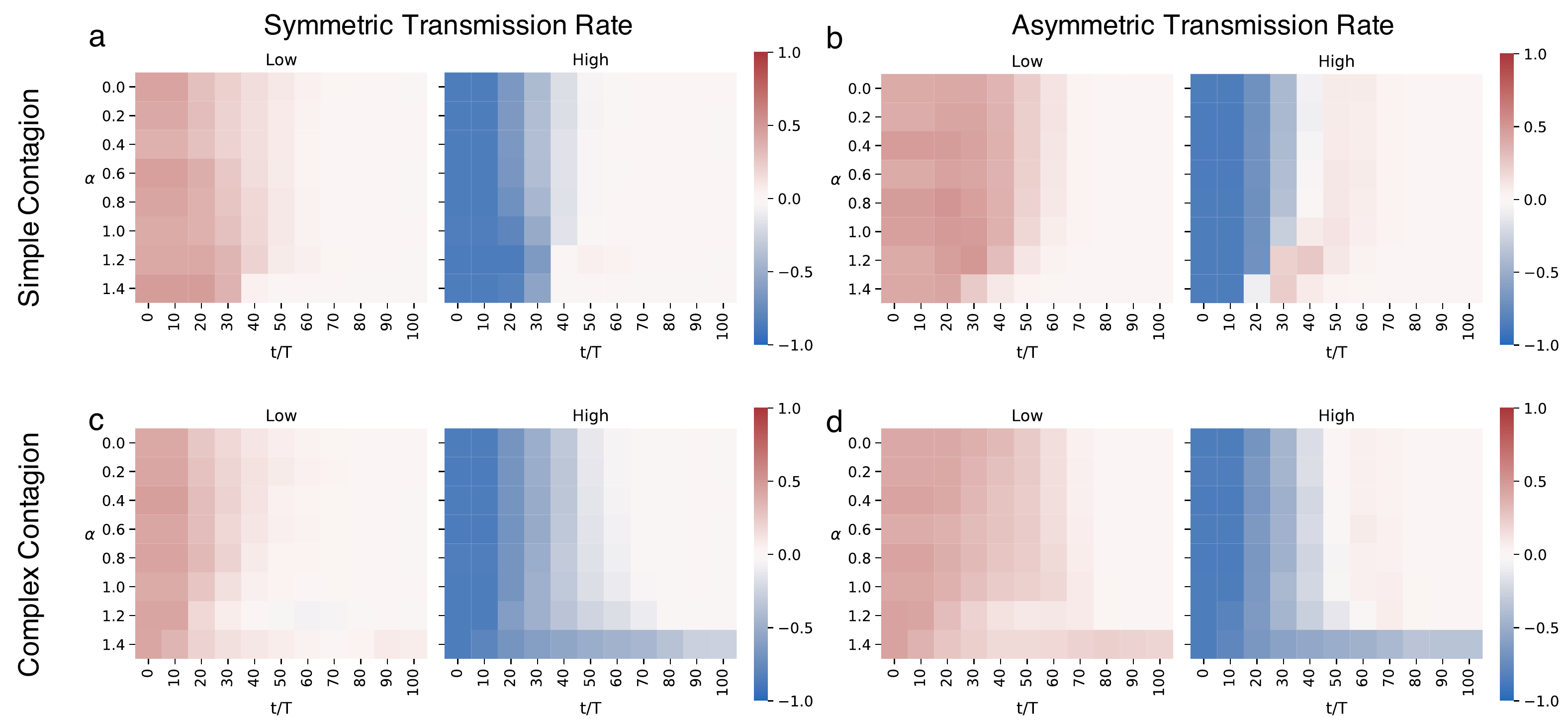}
    \caption{\textbf{Information Spreading Equality for Homophily BA with Different Preferential Attachment Strengths $\alpha$.} We plot the spreading equality for Homophly BA networks with low and high minority seeding portions. Each plot is a heatmap, where the x-axis represents the relative time $t/T$, the y-axis represents the graph's preferential attachment strength $\alpha$, and the color represents $\Delta I(t/T)$. Recall that $\Delta I(t/T) = 0$ represents equality. When $\alpha < 1$, we observe few differences in information equality, implying that for sublinear preferential attachment, other factors such as homophily are more important for information access equality. However, for $\alpha \geq 1$ (which is often the case in real-world networks), the type of contagion is important. For example when $\alpha = 1.4$, simple contagion achieves information access equality quickly (shown in (a) and (b)). This is not true under complex contagion (shown in (c) and (d)).
    This highlights the importance of the type of contagion on the network.}
    \label{fig:homoBA_alpha_info_equality}
\end{figure}

\paragraph{What is the influence of preferential attachment strength on information access equality?}
We experiment with $\alpha$ between 0 and 1.4, from no preferential attachment ($\alpha$ = 0) to  preferential attachment ($\alpha >1$). Fig.~\ref{fig:homoBA_alpha_info_equality} shows the impact of $\alpha$ on information access equality . When $\alpha < 1$, we find limited differences in information equality, implying that for sublinear preferential attachment, other factors such as homophily are more important for information equality. However, for $\alpha \geq 1$, under simple contagion (Fig.~\ref{fig:homoBA_alpha_info_equality}a, b) we observe a slight decrease and then an increase in equality. $\alpha = 1.4$ achieves the most information access equality. However, under complex contagion (Fig.~\ref{fig:homoBA_alpha_info_equality}c, d), we find that $\alpha = 1.4$ achieves the least information access equality. This could be related to the different functions hubs play under simple and complex contagion. Under simple contagion, hubs are pathways, while under complex contagion, hubs are bottlenecks. The observation suggests that the occurrence of hubs on information access equality seems to be significant only under super-linear preferential attachment.

\begin{figure}[ht!]
    \centering
    \includegraphics[scale = 0.38]{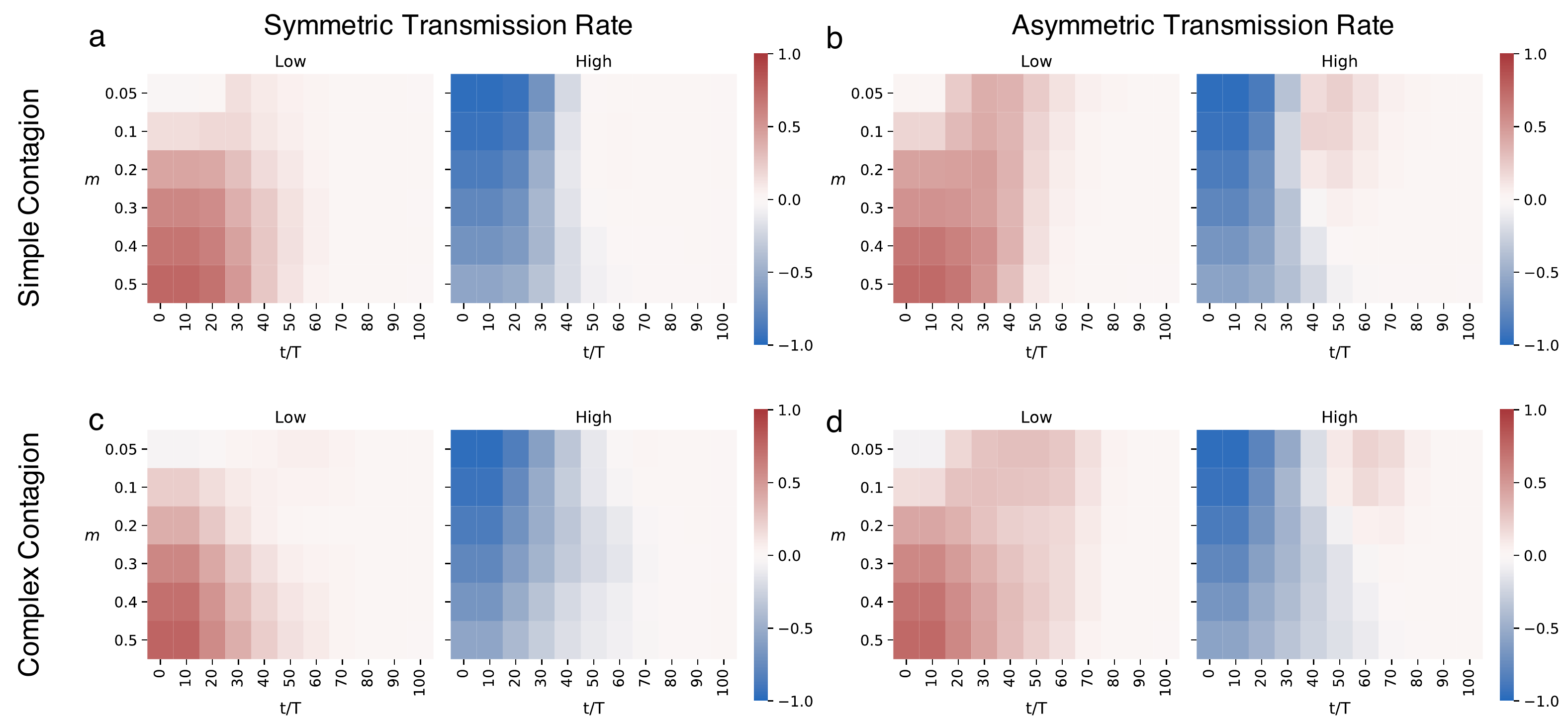}
    \caption{\textbf{Information Spreading Equality for Homophily BA with Different Minority Portions $m$.}  We plot the spreading equality for Homophly BA networks with low and high minority seeding portions. Each plot is a heatmap, where the x-axis represents the relative time $t/T$, the y-axis represents the portion of the nodes that are in the minority group $m$, and the color represents $\Delta I(t/T)$.  Recall that $\Delta I(t/T) = 0$ represents equality. The influence of $m$ on information access equality is strongly dependent on the minority seeding portion and is independent of contagion type and transmission rate. At low minority seeding portion, more equality is achieved with the lowest $m$ (0.05 in our experiments).}
    \label{fig:homoBA_m_info_equality}
\end{figure}

\paragraph{What is the influence of minority portion on information access equality?}
For the minority portion $m$, we experiment in the range between 0.05 (almost no minority nodes) and 0.5 (no population difference between majority and minority nodes). We find that the influence of $m$ on information access equality is strongly dependent on minority seeding portion, independent of contagion type and transmission rate (see Fig.~\ref{fig:homoBA_m_info_equality}).
At low minority seeding portion, more equality is achieved with the lowest $m$ (0.05 in our experiments). 
At high seeding proportion, the higher $m$ values achieve more equality (0.5 under our experiment).
This can be explained by the fact that minority seeding portion (as an absolute value), has a different relative effect among different minority populations. For example, minority seeding portion of 0.3 is considered lower when $m = 0.5$ compared to $m = 0.1$.

\begin{figure}[ht!]
    \centering
    \includegraphics[scale = 0.38]{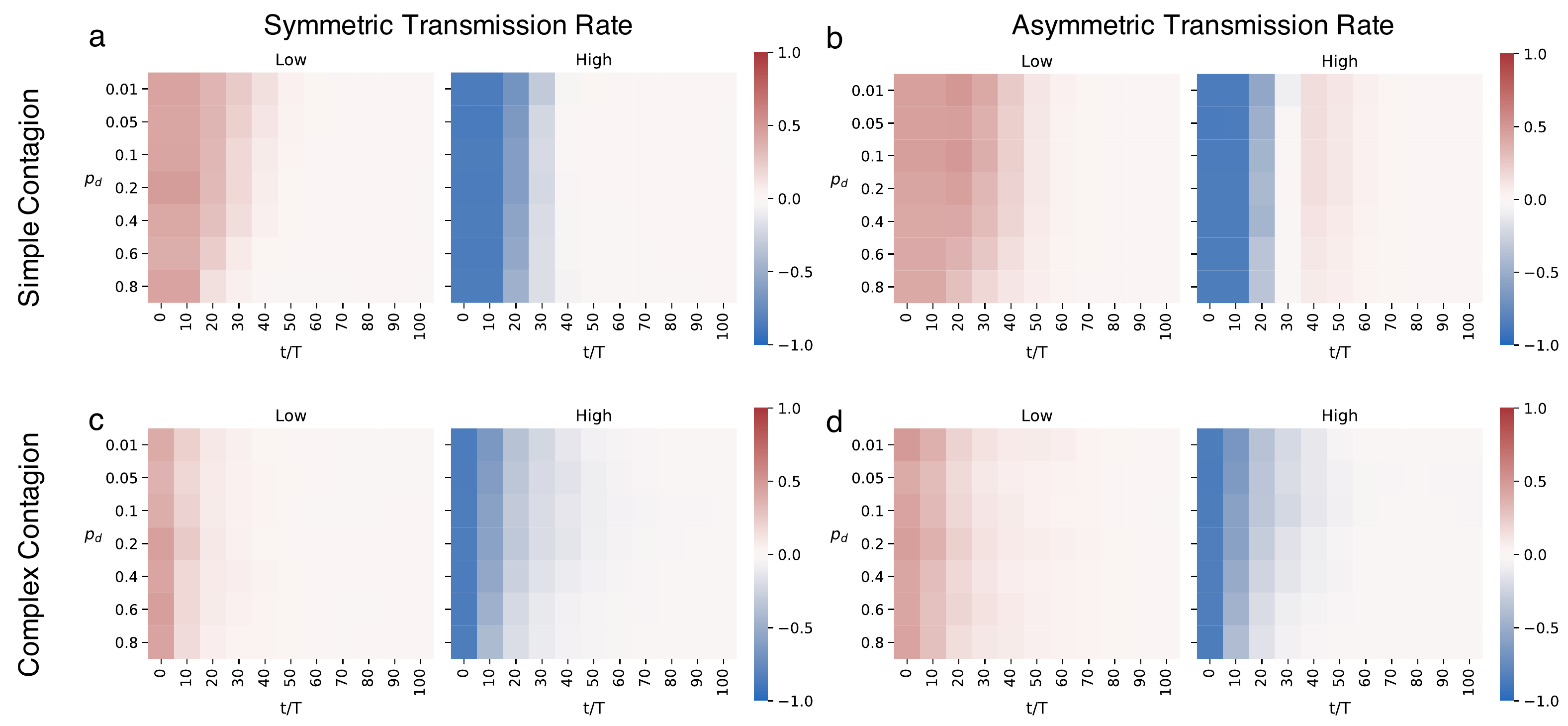}
    \caption{\textbf{Information Spreading Equality for Diversified BA with Different Diversification Probabilities $p_d$.} We plot the spreading equality for Diversified BA  networks with low and high minority seeding portions. Each plot is a heatmap, where the x-axis represents the relative time $t/T$, the y-axis represents the diversification probability $p_d$, and the color represents $\Delta I(t/T)$.  Recall that $\Delta I(t/T) = 0$ represents equality. In general, we observe that the higher the $p_d$, the higher the information access equality, especially for low minority seeding portion. 
    }
    \label{fig:openmindBA_po_info_equality}
\end{figure}

\paragraph{What is the influence of diversity on information access equality?}
We investigate the behavior of Diversified Homophily BA with different $p_d$ values: 0.01, 0.05, 0.1, 0.2, 0.4, 0.6 and 0.8. We set $m = 0.2$, $h =0.8$, $l_d = 1$, and $\alpha = 1$. We see that, in general, higher $p_d$ values achieves higher information access equality, especially for low minority seeding portion (Fig.~\ref{fig:openmindBA_po_info_equality}), which matches our observation that higher $p_d$ has more degree equality and less homophily. 

\subsection*{Experiment 3: Information Access Equality in Real-world Networks}
\begin{figure}[ht!]
    \centering
    \includegraphics[width=.8\columnwidth]{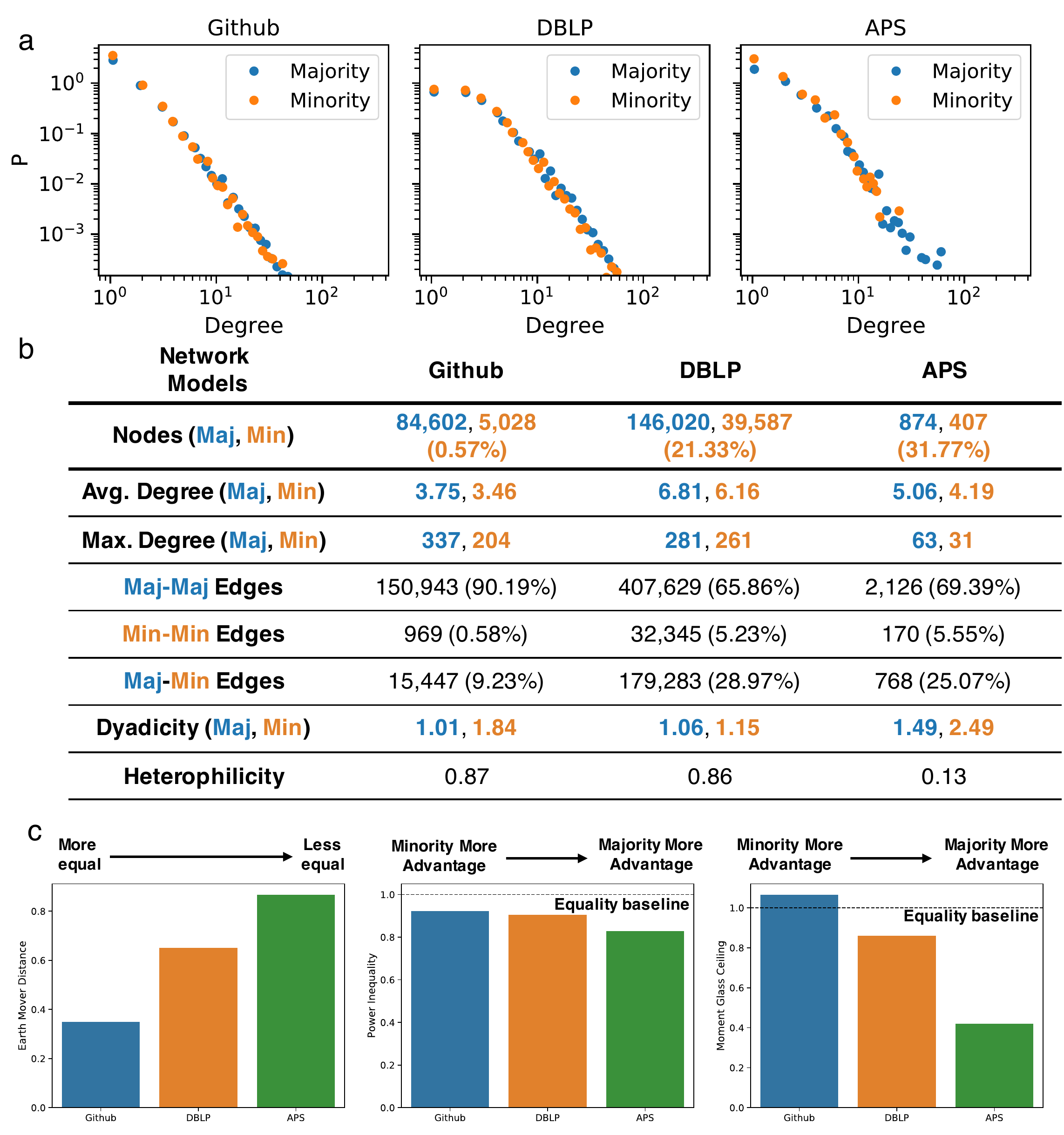}
    \caption{\textbf{Statistics of Real Networks.}
\textbf{(a) Degree Distributions.} We see that all networks have heavy-tail degree distributions, but we note that the minority group in APS has a much shorter tail.
\textbf{(b) Basic Statistics.} The table lists basic statistics, including node composition, average and maximum degree, edge composition, dyadicity and heterophilicity scores for the three networks. Github has the lowest proportion of minority nodes, while APS has the highest proportion of minority nodes. DBLP has the highest average and maximum degree. All the networks have dyadicity greater than 1 and heterophilicity less than 1, indicating homophilic behavior. Among them, APS has the highest dyadicity value and the lowest heterophilicity value.
\textbf{ (c) Social Captial Equality Measures.} The Earth Mover Distance of the three networks are all positive, indicating that the degree distribution is different between the groups. APS has the highest Earth Mover Distance. All networks have power inequality less than 1, indicating that minority groups have a lower average degree. For the moment glass ceiling, GitHub is slightly higher than 1, but the other two networks both have a moment glass ceiling lower than 1, indicating that there are fewer minority nodes as hubs.
    \label{fig:real_network_comparison}}
\end{figure}

We experiment with three real-world networks: (1) the Github follower network, (2) the DBLP collaboration network, and (3) the APS citation network~\cite{lee2019homophily}. The Github and DBLP networks have gender as a grouping attribute. APS has research field as the grouping attribute. We filter the networks by including only labeled nodes and selecting the largest connected component. The basic statistics of the three networks are in Table~\ref{fig:real_network_comparison}b. For the information access simulation, we keep the parameter the same as in Experiments 1 and 2, but change the seed number $s$ to 0.2\% of the number of nodes in each network. For APS, 0.2\% of the nodes is only 2 nodes, which makes it difficult to enforce a different portion of minority seeding, so we set $s = 5$.

Table~\ref{fig:real_network_comparison}b lists the dyadicity and heterophilicity of the real networks. We note that all networks have dyadicity greater than 1 and heterophilicity less than 1, indicating homophily in the network. Comparing across the networks, we observe that APS has the highest dyadicity and the lowest heterophilicity, followed by Github and DBLP.

Figure~\ref{fig:real_network_comparison}(a) shows the degree distributions of of the real networks. Basic statistics for these networks are in (Fig.~\ref{fig:real_network_comparison}b), and social capital (degree) equality measures are in (Fig.~\ref{fig:real_network_comparison}c) . We observe that all networks have heavy-tailed degree distributions, and in terms of degree equality, Github, DBLP, and APS are each from more to less equal.

\begin{figure}[ht!]
    \centering
    \includegraphics[scale = 0.38]{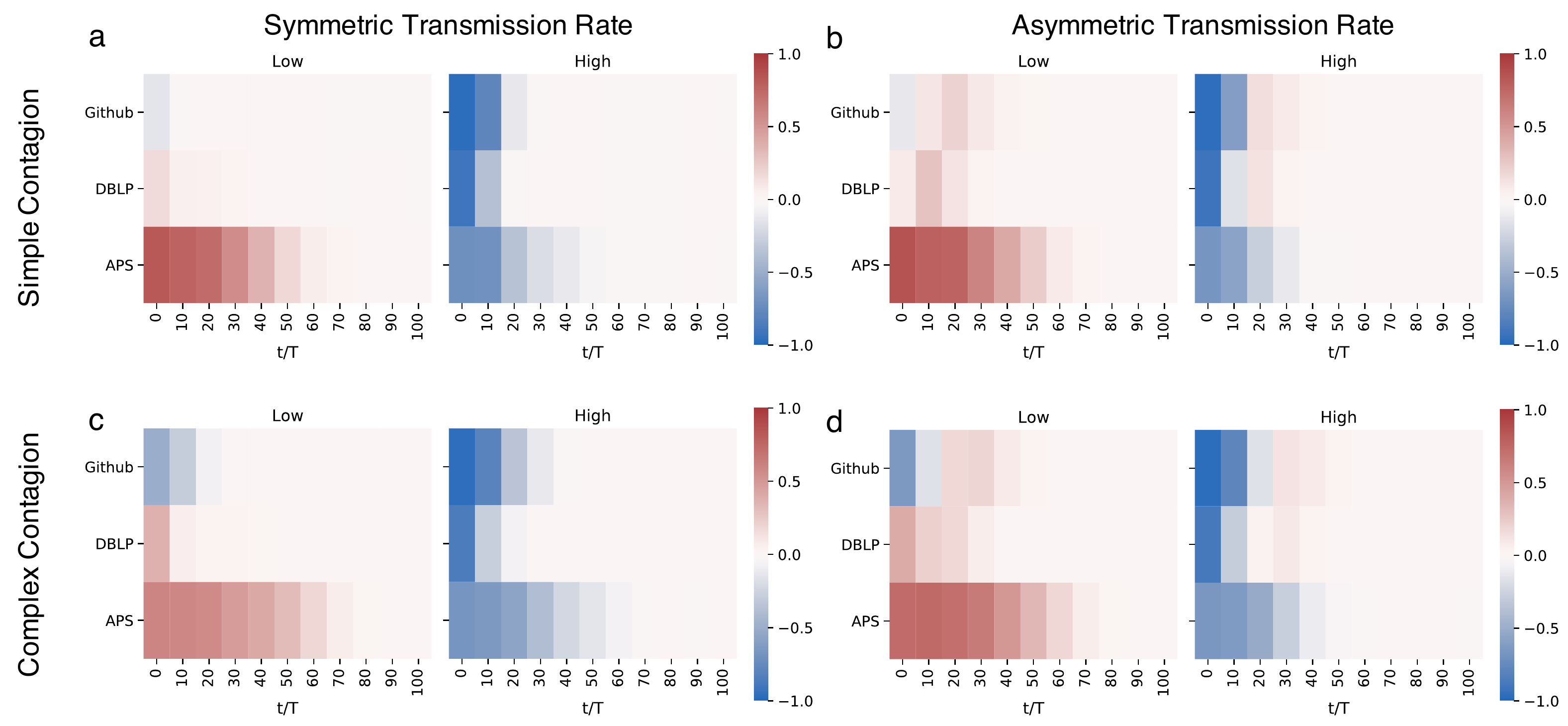}
    \caption{\textbf{Information Spreading Equality of Real Networks.} We plot the spreading equality with low and high minority seeding portions. Each plot is a heatmap, where the x-axis represents the relative time $t/T$, the y-axis represents the different real-world networks, and the color represents $\Delta I(t/T)$. Recall that $\Delta I(t/T) = 0$ represents equality.  For Github, we find that initially $\Delta I(t/T) < 0$, indicating that minority nodes have a greater advantage. This is because Github has the smallest minority portion, and this is consistent with the observation for influence on equality across different $m$ values (see Fig.~\ref{fig:homoBA_m_info_equality}). Across all process settings, we find that APS takes the longest to reach equality, which is consistent with our previous observation that APS is less equal in degree and has a higher homophily level. DBLP is the most equal in information access, which is consistent with lower homophily. We find that the information access equality landscape depends on different process settings. For example, under asymmetric transmission rate, we notice that $\Delta I(t/T)$ becomes positive for Github and DBLP under high seeding portion, which is not the case under symmetric transmission rate. We also find that achieving equality is much harder under complex contagion and asymmetric transmission rate.
    }
    \label{fig:real_network_info_equality}
\end{figure}

Figure~\ref{fig:real_network_info_equality} shows information access equality on real networks. For Github, we find that initially $\Delta I(t/T) < 0$, indicating that nodes that are in the minority have a greater advantage. This is because Github has the smallest minority portion, and this is consistent with the observation for influence on equality with different $m$ (see Fig.~\ref{fig:homoBA_m_info_equality}). For all different process settings, we find that APS takes the longest to reach equality, which is consistent with our previous observation that APS is less equal in degree and has higher homophily level. DBLP is most equal in information access, which is consistent with lower homophily. We find that the information access equality landscape depends on different process settings. For example, under asymmetric transmission rate, we notice that $\Delta I(t/T)$ becomes positive for Github and DBLP under high seeding portion, which is not the case under symmetric transmission rate. We also find that achieving equality is much harder under complex contagion and asymmetric transmission rate. These variations are also consistent with our earlier observation that information access is not only related to the network but also to the process setting (see Fig.~\ref{fig:real_network_info_equality}).

\subsection*{Experiment 4: Relationship Between Information Equality and Spreading Efficiency}
Research on fairness in AI has shown a phenomenon called the ``price of fairness''~\cite{menon2018cost,corbett2017algorithmic}, where researchers have found that fairness comes at a cost to other performance measures of interest. For information spreading, the relationship between equality and efficiency is a rather important one, especially when the information is time-sensitive. Past work has found the ``price of fairness'' under information maximization~\cite{tsang2019group}. 

To measure spreading efficiency, we inspect the value of $I(t)$ in the spreading process. Since comparing efficiency among various spreading processes is affected by network statistics such as the size of the network, the average degree, etc., we examine spreading efficiency for the networks created for Experiment 1.

\begin{figure}[ht!]
    \centering
    \includegraphics[scale = 0.38]{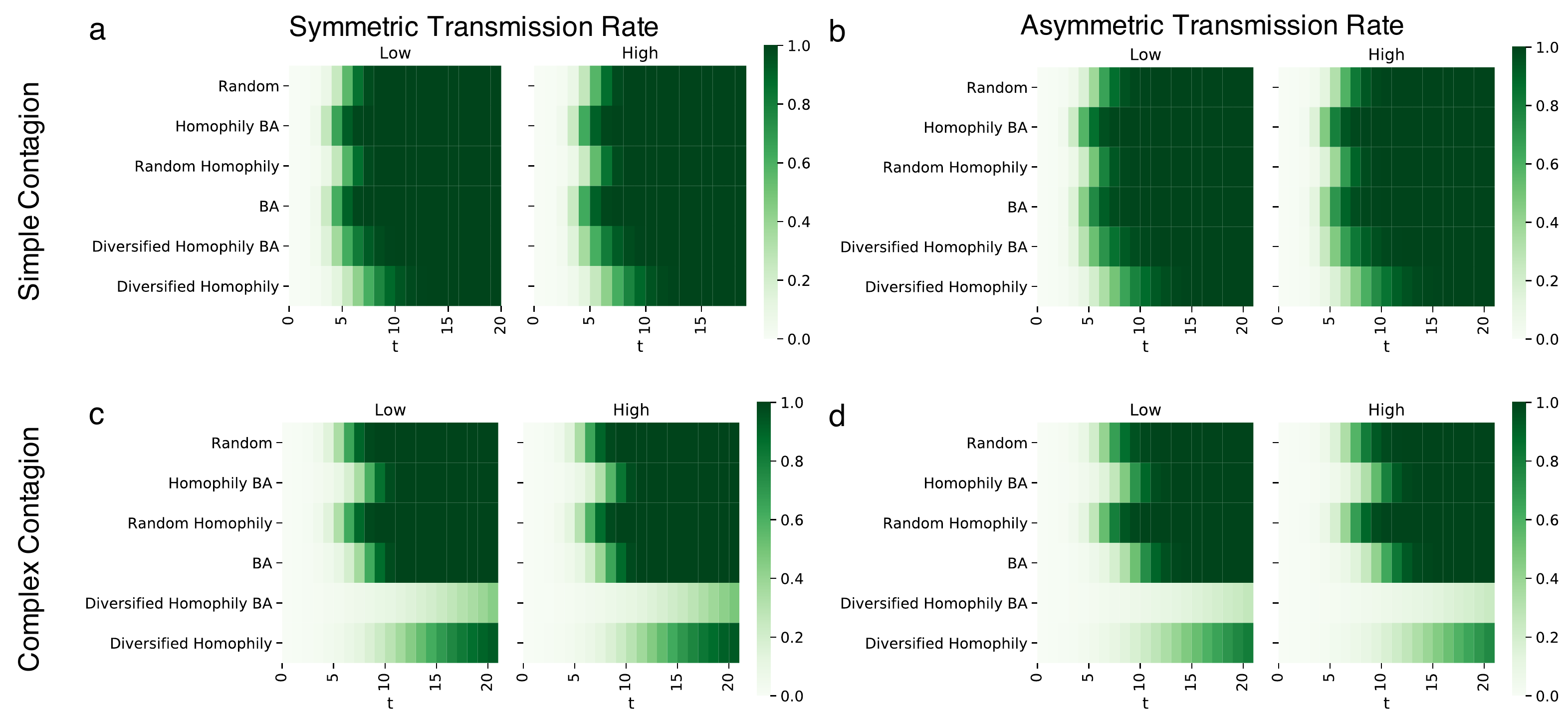}
    \caption{\textbf{Information Spreading Efficiency of Different Generative Models of Complex Networks.} Similar to the information access equality, we also inspect information access efficiency as a heatmap. Here the x-axis is the actual time $t$, the y-axis is the different models, and the color represents the fraction of nodes are in state I (i.e., infected by the information) at a given time. We notice that the driving force of the spreading efficiency is the type of contagion. For the case of simple contagion (shown in (a) and (b)), we observe that Homophily BA and BA spread much faster, while Diversified Homophily is the slowest. In the case of complex contagion (shown in (c) and (d)), we observe that Random Network and Random Homophily are the most efficient, followed by BA and Homophily BA. We conjecture that this is due to the fact that hubs become bottlenecks in complex contagion. Diversified Homophily and Diversified Homophily BA are much slower in spreading, indicating the trade-off between equality and efficiency of information access.}
        \label{fig:network_info_scale}
\end{figure}

Figure~\ref{fig:network_info_scale} shows the information spreading efficiency of different network models. Similar to equality, we plot the spreading process as a heatmap, where the x-axis represents the actual time $t$, the y-axis represents the different models, and the color represents the fraction of already infected nodes $I(t)$. We notice that the driving force for spreading efficiency is the type of contagion. In simple contagion (Fig.~\ref{fig:network_info_scale}a, b), Homophily BA and BA are much more efficient at spreading, while Diversified Homophily is the least efficient. In complex contagion (Fig.~\ref{fig:network_info_scale}c, d), we observe that Random Network and Random Homophily are the most efficient, followed by BA and Homophily BA, possibly due to the fact that hubs are not beneficial in complex contagion. Diversified Homophily BA and its variant are still the least efficient in spreading. Recall that Diversified Homophily BA and its variation achieve more information access equality, indicating that inter-group edges may lead to a trade-off between equality and efficiency in information spreading.

By inspecting different network parameters, we observe that except for the minority population $m$, all other parameters can affect the information spreading efficiency. For homophily $h$, $h = 1$ always has the lowest efficiency, possibly due to the extreme case where only one edge connects the two groups. We observe that for asymmetric transmission rate, a lower $h$ has a slightly slower spreading speed due to more inter-group edges, again showing the equality and efficiency trade-off of inter-group edges.

We observe a clear trend of the spreading efficiency and $\alpha$. Under simple contagion, the higher the $\alpha$, the faster the spreading. In contrast, under complex contagion, the higher the $\alpha$, the slower the spreading speed. This is to be expected because the emergence of hubs promotes simple contagion but hinders complex contagion. Recall that we see that higher $\alpha$ has higher equality in simple contagion and lower equality in complex contagion, this shows that hubs do not create a trade-off between efficiency and equality.

We observe a slight decrease in spreading efficiency as $p_d$ varies under simple contagion. Under complex contagion, higher $p_d$ significantly decreases the spreading speed, again showing the trade-off between efficiency and equality with increasing inter-group edges.

In short, we find that there is a trade-off between information access equality and information spreading efficiency. This is mainly due to the inter-group edges. However, similar to the observation that equality is related to process setting, this trade-off is also dependent on process setting.

\section{Conclusion and Discussion}
In this paper, we focused on information access equality in complex networks when the population is divided into two mutually exclusive groups: majority vs.~minority. We measured the information access equality of various processes on different network models with different parameters and on three real networks. We find that information access equality depends not only on the network structure, but also on the spreading process, different contagion types (i.e., simple vs.~complex), different transmission rates (symmetric vs.~asymmetric), and different minority proportions. Although it is very difficult to draw a single conclusion about what type of networks can promote information access equality, we find that, in general, more inter-group edges can help achieve equality. However, we note one drawback to more inter-group edges: they may reduce the efficiency of information spreading. Designing a network with information access equality requires more knowledge about the spreading process itself. Our findings can be used to guide recommendations for mechanistic design of social networks that foster more equal information access. For example, social networking platforms, such as Facebook and LinkedIn, can recommend new connections the lead to more equal information access for various spreading processes on their platforms. Deployment of our findings in a real-world system is part of our future work. 

\bibliographystyle{unsrt}  
\bibliography{fairnet}

\end{document}